\documentclass[prl,twocolumn,epsfig,rotate,superscriptaddress,showpacs]{revtex4}

\usepackage{latexsym} 
\usepackage{amssymb}  
\usepackage{amsfonts} 
\usepackage{amsbsy}   
\usepackage{amsmath} 
\usepackage{graphicx}       
\usepackage{multirow}  
\usepackage[dvips]{color}
\usepackage{bm}   

\newcommand{\<}{\langle}
\renewcommand{\>}{\rangle} 
\newcommand{\txt}{\textstyle}

\newcommand\eqn[1]{(\ref{#1})}      
\newcommand\Eqn[1]{Eq.~(\ref{#1})}  
\newcommand{\e}{{\rm e}}   

\newcommand{\nn}{\nonumber}

\newcommand{\half} {{\txt \frac{1}{2}}}

\newcommand{\cP}{\ensuremath{\mathcal{P}}}
\newcommand{\cT}{\ensuremath{\mathcal{T}}}

\makeatletter 

\begin{document}
\title{\bf Geometric Phase in \cP\cT-Symmetric Quantum Mechanics}
\author{Jiangbin Gong}
\affiliation{Department of Physics, National University of Singapore,
Singapore, 117542}
\affiliation{ Centre for Computational Science and Engineering,
National University of Singapore, Singapore, 117542}
\affiliation{NUS Graduate School for Integrative Sciences and Engineering, Singapore, 117597}
\author{Qing-hai Wang}
\affiliation{Department of Physics, National University of Singapore,
Singapore, 117542}

\date{\today}

\begin{abstract}
Unitary evolution in \cP\cT-symmetric quantum mechanics with a time-dependent metric is found to yield
a new class of adiabatic processes.
As an explicit example, a Berry-like phase associated with a \cP\cT-symmetric two-level system is derived and interpreted as the
 flux of a fictitious monopole with a tunable charge plus a singular string component with non-trivial phase contributions. The Hermitian analog of our results is also discussed.
\end{abstract}

\begin{titlepage}
\maketitle
\renewcommand{\thepage}{}          
\end{titlepage}

Parity-time-reversal (\cP\cT) symmetric  quantum mechanics (QM) has emerged as a complex extension of conventional QM, via
non-Hermitian Hamiltonians that possess unbroken \cP\cT-symmetry and hence real energy spectrum \cite{BB}.
On a deep level \cP\cT-symmetric QM is not expected to challenge the current quantum theory.
Nonetheless, some fascinating features of \cP\cT-symmetric QM have been revealed \cite{Review}, with their implications being
pursued both experimentally \cite{exp} and theoretically \cite{theory}.
It is now certain that \cP\cT-symmetric QM will bring more insights and results that are not obvious in conventional QM.

In \cP\cT-symmetric QM, a given non-Hermitian Hamiltonian $H({\bf X})$, here parameterized by a multi-dimensional vector
${\bf X}$, is self-adjoint with respect to a generalized inner product. In particular,
a Hermitian, positive-definite metric operator $W({\bf X})$ constructed from
the \cP\cT\ symmetry of a system defines a new inner product, denoted $(\Psi,\Phi)_{W}$  for two arbitrary states $\Psi$ and $\Phi$. In Dirac's bra and ket notation,
$(\Psi,\Phi)_{W}\equiv \langle \Psi|W({\bf X)}|\Phi\rangle$.  The self-adjoint condition of $H({\bf X})$ then becomes
\begin{eqnarray}
W({\bf X})H({\bf X})&=&H^{\dagger}({\bf X})W({\bf X}),
\label{eqn:WH}
\end{eqnarray}
which yields real spectrum and orthonormal eigenstates.

Because the metric operator $W({\bf X})$ must be always attached to $H({\bf X})$, a time-dependent $H({\bf X})$ via ${\bf X}={\bf X} (t)$ calls for a time-dependent metric, an interesting situation previously overlooked.  This motivates us to exploit time-dependent problems in \cP\cT-symmetric QM to explore fundamental issues, among which adiabatic manipulation
 and Berry-phase related problems are
of broad interest \cite{Berry,book}. Indeed, to preserve the  $W({\bf X})$-based inner product and hence unitarity,
the time-evolution equation in \cP\cT-symmetric cases is shown to have special features, thereby defining new dynamical problems from the very beginning.

Specifically, we present below (i) general considerations of the adiabatic time evolution associated with a time-dependent $H[{\bf X}(t)]$ in \cP\cT-symmetric QM, and (ii) intriguing geometric phase results based on a six-parameter complete characterization of $2\times 2$ \cP\cT-symmetric Hamiltonian matrices \cite{qh}. The found Berry-like phase in a representative case is interpreted as the flux of a fictitious monopole with a tunable charge, plus a singular component similar to a Dirac string but with a non-trivial effect. To gain more insights, we also analyze the Hermitian analog of our non-Hermitian problem, finding that the context of time-dependent \cP\cT-symmetric QM also opens up an interesting class of geometric-phase problems in conventional QM.  Furthermore, we discuss common features shared by this work and a recent study regarding geometric phases in nonlinear systems.

Consider then the following linear evolution equation for \cP\cT-symmetric QM (setting $\hbar=1$ throughout),
\begin{equation}
i \frac{d}{dt} |\Psi(t)\> = \Lambda(t) |\Psi(t)\>.
\label{eqn:Schrodinger}
\end{equation}
The general form of $\Lambda(t)$ can be determined by the unitarity condition. That is, for two arbitrary solutions $|\Phi(t)\>$ and $|\Psi(t)\rangle$,
\begin{equation}
\frac{d}{dt}\<\Phi(t)|W[{\bf X}(t)]|\Psi(t)\> = 0.
\label{eqn:unitarity}
\end{equation}
Equations (\ref{eqn:Schrodinger}) and (\ref{eqn:unitarity}) directly lead to
\begin{equation}
i \frac{dW[{\bf X}(t)]}{dt} =\Lambda^\dag(t) W[{\bf X}(t)] - W[{\bf X}(t)]\Lambda(t).
\end{equation}
Letting
\begin{equation}
\Lambda(t) = H[{\bf X}(t)] -i K[{\bf X}(t)],
\end{equation}
where $H[{\bf X}(t)]$ is the time-dependent \cP\cT-symmetric Hamiltonian satisfying \Eqn{eqn:WH}, one arrives at the following condition for $K[{\bf X}(t)]$:
\begin{equation}
\bm{\nabla} W[{\bf X}(t)]\cdot \frac{d{\bf X}}{dt} =K^{\dag}[{\bf X}(t)] W[{\bf X}(t)] + W[{\bf X}(t)]K[{\bf X}(t)],
\label{delta-eq}
\end{equation}
where $\bm\nabla$ is the gradient in the ${\bf X}$-space.
 Several remarks on \Eqn{delta-eq} are in order.  First, if $K[{\bf X}(t)]$ satisfies \Eqn{delta-eq}, so does $K[{\bf X}(t)]+i \tilde{H}(t)$, where $\tilde{H}(t)$ is an arbitrary \cP\cT-symmetric Hamiltonian with $W[{\bf X}(t)]\tilde{H}(t)=\tilde{H}^{\dagger}(t)W[{\bf X}(t)]$. This arbitrariness hence just reflects a different partition of $\Lambda(t)$.  Second, in general $K[{\bf X}(t)]$ is a function of $d{\bf X}/dt$, the velocity in the  ${\bf X}$-space.  Third, to find an interesting sample solution to \Eqn{delta-eq}, we put forward
 below an additional condition  $K[{\bf X}(t)]=K^{\dag}[{\bf X}(t)]$. 
 With these clarified, one can now adopt a solution  $K[{\bf X}(t)]= {\bf M}[{\bf X}(t)]\cdot (d{\bf X}/dt)$, with
\begin{equation}
\bm{\nabla} W({\bf X}) ={\bf M}({\bf X}) W({\bf X}) + W({\bf X}){\bf M}({\bf X})
\label{M-eq2}
\end{equation}
and ${\bf M}({\bf X})={\bf M}^{\dag}({\bf X})$.
Clearly, if $d{\bf X}/dt\rightarrow 0$, then the constructed $K[{\bf X}(t)]\rightarrow 0$.

Next we examine the time dependence of $|\Psi(t)\>$ [see \Eqn{eqn:Schrodinger}] for a slowly-varying ${\bf X}$.  Because in the adiabatic limit, $\Lambda(t)=H[{\bf X}(t)]$, the natural adiabatic representation should comprise instantaneous eigenstates of $H[{\bf X}(t)]$, denoted $|\psi_n[{\bf X}(t)]\>$.  Specifically, if
\begin{equation}
H[{\bf X}(t)] \left|\psi_n[{\bf X}(t)]\right\> = E_n[{\bf X}(t)] \left|\psi_n[{\bf X}(t)]\right\>,
\label{eqn:eigen}
\end{equation}
with the orthonormal condition
$\<\psi_n[{\bf X}(t)]|W[{\bf X}(t)]|\psi_m[{\bf X}(t)]\>=\delta_{mn}$, then we can expand $|\Psi(t)\>$ as
\begin{equation}
|\Psi(t)\> = \sum_n c_n(t) \e^{i\alpha_n(t)} |\psi_n[{\bf X}(t)]\>,
\label{expansion}
\end{equation}
where $c_n$ is the expansion coefficient, and $\alpha_n$ is a dynamical phase defined as the time-integral of the instantaneous real energy eigenvalue $E_n[{\bf X}(t)]$, i.e.,
$\alpha_n(t) \equiv -  \int^t_0 E_n[{\bf X}(t')]d t' $.
Substituting \Eqn{expansion} into \Eqn{eqn:Schrodinger}, one obtains
\begin{eqnarray}
\dot{c}_m(t) = - c_m(t){\cal  G}_{m}(t) - \sum_{n\neq m} c_n(t) \e^{i[\alpha_n(t)-\alpha_m(t)]} {\cal T}_{mn},
\label{cmeq}
\end{eqnarray}
where
\begin{eqnarray}
{\cal  G}_{m}(t)&\equiv& \biggl\{\<\psi_m({\bf X})|W({\bf X})|\bm\nabla \psi_m({\bf X})\> \nn \\
&& +\ \<\psi_m({\bf X})|W({\bf X}){\bf M}({\bf X})|\psi_m({\bf X})\> \biggr\} \cdot \frac{d{\bf X}}{dt}, \\
{\cal  T}_{mn}(t)&\equiv& \biggl\{\<\psi_m({\bf X})|W({\bf X})\left[\frac{\bm\nabla H({\bf X})}{E_n({\bf X})-E_m({\bf X})}\right]|\psi_n({\bf X})\> \nn\\
&&+ \<\psi_m({\bf X})|W({\bf X}){\bf M}({\bf X})]|\psi_n({\bf X})\> \biggr\}\cdot \frac{d{\bf X}}{dt}
\end{eqnarray}   with ${\bf X}={\bf X}(t)$.  Clearly, due to the oscillating phase factor  $\e^{i[\alpha_n(t)-\alpha_m(t)]}$ in \Eqn{cmeq}, the transition probabilities between non-degenerate eigenstates $|\psi_n({\bf X})\>$ and $|\psi_m({\bf X})\>$ are negligible
if we have $\bm\nabla H({\bf X}) \cdot(d{\bf X}/dt)/[E_n({\bf X})-E_m({\bf X})]^2\ll 1$ and ${\bf M}({\bf X})\cdot (d{\bf X}/dt)/[E_n({\bf X})-E_m({\bf X})]\ll 1$ (a precise adiabatic condition is out of the scope of this work and the analogous topic in conventional QM is still under study). As such, in the limit of $d{\bf X}/dt\rightarrow 0$, adiabatic evolution in \cP\cT-symmetric QM does exist even for a time-dependent metric, insofar as the time evolving state  can remain to be one instantaneous eigenstate of a time-evolving Hamiltonian.

For adiabatic evolution, the ${\cal G}_{m}$ term in \Eqn{cmeq} gives the solution
$c_{m}(t)=c_m(0)\e^{i\gamma(t)}$, with
\begin{eqnarray}
 \gamma(t) &=& i\int_{{\bf X}(0)}^{{\bf X}(t)}\left[ \<\psi_m({\bf X})|W({\bf X})|\bm{ \nabla}\psi_m({\bf X})\>\right. \nn\\
&&+\ \left. \<\psi_m({\bf X})|W({\bf X}){\bf M}({\bf X})|\psi_m({\bf X})\>\right] \cdot d{\bf X}.
\end{eqnarray}
Because $\gamma(t)$ is solely determined by the geometry of a navigation path in the ${\bf X}$-space and does not depend on the duration of the adiabatic process, it is of a geometric origin. In particular, for a cyclic path, this geometric phase becomes a Berry-like phase
$\gamma_{B}$, i.e.,
\begin{eqnarray}
\gamma_{B}&=& i \oint \left[ \<\psi_m({\bf X})|W({\bf X})|\bm{ \nabla}\psi_m({\bf X})\> \right. \nn\\
&&+\ \left.\<\psi_m({\bf X})|W({\bf X}){\bf M}({\bf X})|\psi_m({\bf X})\>\right] \cdot d{\bf X}.
\label{gammaex}
\end{eqnarray} In \Eqn{gammaex}, the first term is parallel to a familiar Berry-connection integral \cite{Berry}, but the second term represents effects due to a time-dependent metric. It is the sum of these two terms that can ensure a real $\gamma_{B}$.  Note also that  $\gamma_{B}$ is invariant upon the gauge transformation $|\psi_m({\bf X})\>\rightarrow \e^{i f({\bf X})}|\psi_m({\bf X})\>$, where $f({\bf X})$ is an arbitrary continuous phase function.

Let us now apply these general considerations to a two-dimensional Hilbert space. Here, the \cT\ operator is taken as the Dirac conjugate, and a nontrivial $2\times 2$ \cP\ operator is chosen as
\begin{equation}
\cP_{2\times 2}=\vec{e}_r\cdot \vec{\sigma}=\left(
\begin{array}{cc}
\cos\theta & \sin\theta~ \e^{-i\varphi}\\
\sin\theta~ \e^{i\varphi} & -\cos\theta
\end{array}
\right),
\label{eqn:P_2x2}
\end{equation}
where $\vec{{\sigma}}\equiv (\sigma_x,\sigma_y,\sigma_z)$ represents the standard Pauli matrices, and $\vec{e}_r\equiv (\sin\theta\cos\varphi,\sin\theta\sin\varphi,\cos\theta)$. To avoid confusion with the vectors in the ${\bf X}$-space of arbitrary dimension, below we always use $\vec{A}$ to represent a three-dimensional vector $A$.
The general form of \cP\cT-symmetric $2\times 2$ Hamiltonian matrices can then be specified by six real parameters [${\bf X}=(\varepsilon, a,b,\theta,\phi,\delta)]$ \cite{qh}:
\begin{eqnarray}
H_{2\times 2}({\bf X})=\varepsilon + \left(a \vec{e}_r + i b\cos\delta \vec{e}_{\theta} + i b\sin\delta \vec{e}_\varphi\right)\cdot\vec{\sigma},
\label{eqn:H}
\end{eqnarray}
where
$\vec{e}_\theta \equiv (\cos\theta\cos\varphi,\cos\theta\sin\varphi,-\sin\theta)$ and $
\vec{e}_\varphi \equiv (-\sin\varphi,\cos\varphi,0)$. Here we apply
 the constraint $a^2>b^2$ to ensure a non-degenerate real spectrum (see explicit eigenvalue expressions below). The special case of $b=0$ recovers Hermitian matrices with four real parameters. Attached to such a complete characterization of $H_{2\times 2}({\bf X})$ is the following $2\times 2$ metric operator,
\begin{equation}
W_{2\times 2}({\bf X}) = \frac{|a|}{\sqrt{a^2-b^2}} \left(1 + \vec{{\beta}}\cdot\vec{{\sigma}} \right),
\label{w22}
\end{equation}
where
\begin{eqnarray}
\vec{{\beta}} \equiv \frac{b}{a}\left( \sin\delta\, \vec{e}_\theta - \cos\delta\, \vec{e}_\varphi\right).
\end{eqnarray}
The real eigenvalues of $H_{2\times 2}({\bf X})$ are $E_{\pm}=\varepsilon\pm \sqrt{a^2-b^2}$, with eigenvectors
\begin{equation}
|\psi_\pm\> = \frac{1}{\cal N_{\pm}} \left[
\begin{array}{c}
\e^{-i\phi}(a\sin\theta +i b\cos\delta\cos\theta+b\sin\delta)\\
-a\cos\theta+ib\cos\delta\sin\theta\pm \sqrt{a^2-b^2}
\end{array}\right].
\label{eqn:eigenstates}
\end{equation}
Here the normalization factor ${\cal N}_{\pm}$ is determined by ${\cal N}^2_{\pm}=2|a|\sqrt{a^2-b^2}\left[1+(b/a)\sin\delta\sin\theta \mp{\cal U}\cos\theta\right]$, where
\begin{eqnarray}
{\cal U} &\equiv& \frac{\sqrt{a^2-b^2}}{a}.
\end{eqnarray}

The fact that $H_{2\times 2}({\bf X})$ has two more free parameters than a $2\times 2$ Hermitian matrix is stimulating.
In the following we focus on those navigation paths that only involve changes in $\theta$ and $\phi$. Extension to other cases (e.g., change in $\delta$) is certainly worthwhile but straightforward.  In the $(\theta,\phi)$ manifold, $a$, $b$, and $|\vec{{\beta}}|$ are constants and we obtain a simple Hermitian solution to \Eqn{M-eq2}:
\begin{equation}
{\bf M}_{2\times 2}({\bf X}) = \half \bm{\nabla} \vec{{ \beta}}({\bf X})\cdot\vec{{\sigma}}.
\label{m22}
\end{equation}
Substituting Eqs.~\eqn{w22}, \eqn{eqn:eigenstates}, and \eqn{m22} into \Eqn{gammaex} leads us to a lengthy calculation, but we finally arrive at a rather compact expression for the geometric phase.
For example, associated with $|\psi_{+}\>$, the Berry-like phase denoted $\gamma_{B}^{+}$ is found to be
\begin{eqnarray}
\gamma_{B}^{+}= \oint \left[ F_{\phi}({\bf X}) d\phi + F_{\theta}({\bf X})d\theta\right],
\label{gammaB1}
\end{eqnarray}
where
\begin{eqnarray}
F_{\phi}({\bf X})& = & \half\left(1+ {\cal U} \cos\theta\right),
\label{Fphi} \\
F_{\theta}({\bf X})& =& \frac{(b/a)\cos\delta}{2[{1+(b/a)\sin\delta\sin\theta -{\cal U}\cos\theta}]}.
\end{eqnarray}

To better understand $\gamma_{B}^{+}$ we now treat $\theta$ and $\phi$ as two spherical angular coordinates on a sphere defined by $\vec{r}\equiv r\vec{e}_{r}\equiv (x,y,z)$.  If we define
\begin{eqnarray}
\vec{A} &\equiv& \frac{F_{\theta}({\bf X})}{r}\vec{e}_{\theta} + \frac{F_{\phi}({\bf X})}{r\sin\theta}\vec{e}_{\phi},
\label{vectorA}
\end{eqnarray}
then because $F_{\phi}({\bf X}) d\phi + F_{\theta}({\bf X})d\theta= \vec{A}\cdot d\vec{r}$, \Eqn{gammaB1} becomes
\begin{eqnarray}
\gamma_{B}^{+}&=&\oint \vec{ A}\cdot d\vec{r}.
\label{linein}
\end{eqnarray}
This motivates us to regard $\vec{A}$ defined above as a three-dimensional vector potential in the $\vec{r}$-space.
If the surface enclosed by a closed path on the sphere does not include $\theta=0,\pi$, i.e., the north or south pole, then $\vec{A}$ is well behaved everywhere on this surface, and as a result Stokes' theorem converts \Eqn{linein} into a surface integral: $\gamma_{B}^{+}=\int\int \vec{\nabla}\times \vec{ A}\cdot d\vec{S}$. Further using $\partial F_{\theta}({\bf X})/\partial \phi =0$ and the standard curl operator in spherical coordinates, one finds $\gamma_{B}^{+}=-({\cal U}/2)\int\int \sin\theta d\theta d\phi=-({\cal U}/2)\Omega$, where ${\Omega}$ is the solid angle spanned by the closed path.
On the other hand, if a closed path does enclose the north pole of the sphere (for convenience we always assume a counter-clockwise path), then the circulation integral in \Eqn{linein} can be converted into two integrals: one is on a deformed closed path circumventing the north pole and the other is on an infinitely small circle around the north pole.
In this second situation, we obtain
$\gamma_{B}^{+}=-({\cal U}/2)\Omega+ (1+{\cal U})\pi$.  The sole contribution by the north (or south) pole can then be found by setting $\Omega=0$ (or $\Omega=4\pi$). This is consistent with a direct calculation from  \Eqn{gammaB1} by fixing $\theta$ at $0$ (or $\pi$).

With a fictitious magnetic field denoted $\vec{ B}_{\text{eff}}$, $\gamma_{B}^{+}$ for an arbitrary closed path on the sphere can now  be expressed as a magnetic flux through a surface enclosed by the closed path, i.e., $\gamma_{B}^{+}=\int\int \vec{ B}_{\text{eff}}\cdot d\vec{S}$, where
\begin{eqnarray}
\vec{B}_{\text{eff}} &=& \left(1+{\cal U}\right)\pi\,\delta(x)\delta(y) \vec{e}_{z} - \frac{\cal U}{2}\frac{\vec{r}}{r^3}
\label{central}
\end{eqnarray}
with $\vec{e}_{z}\equiv\vec{e}_{r}(\theta=0)$.
Equation (\ref{central}) is our main result for a Berry-like phase of a \cP\cT-symmetric two-level system. $\vec{B}_{\text{eff}}$ represents a fictitious singular field pointing at the north pole, plus a virtual magnetic monopole. The magnitude of the charge of such a virtual monopole is {\it continuously  tunable}: it is given by a unit charge $g_{0}=-\half$ multiplied by the factor ${\cal U}\in(0,1]$ for $a>0$ or  ${\cal U}\in[-1,0)$ for $a<0$. The singular $\vec{e}_{z}$ component of Eq.~(\ref{central}) is analogous to the so-called Dirac string in the Dirac quantization rule for the product of a magnetic charge and an electric charge.
Remarkably, the contribution by the singular string component here is seen to be physically relevant: it generates a flux in the range of
$[0,\pi)$ or $(\pi,2\pi]$.  Note that, in the limit of $b=0$ and hence ${\cal U}=\pm 1$, the monopole field in \Eqn{central} carries a charge $\pm g_0$. Under the same limit
the first component of \Eqn{central} reduces to the trivial and unobservable Dirac string because it can only contribute a phase of $0$ or $2\pi$. This makes it clear that the standard results for the Berry phase of a Hermitian two-level system can be all recovered for $b=0$.

The issue of geometric phase in non-Hermitian systems was first touched upon in a very interesting study in Ref.~\cite{wright}.  However, the spectrum studied therein is not real. In addition, the construction of a geometric phase in Ref.~\cite{wright} was based upon a time-dependent Schr\"{o}dinger equation and its non-equivalent adjoint. By contrast, here the spectrum is always real, the adiabatic condition is similar to that in the Hermitian case, and the evolution of any state in the Hilbert space satisfies a common dynamical equation.

To shed more light on our results we next examine a Hermitian analog of our geometric-phase problem. Indeed,
the real spectrum of a \cP\cT-symmetric Hamiltonian $H_{2\times 2}({\bf X})$ implies the existence of a Hermitian Hamiltonian $h({\bf X})$ with the same spectrum \cite{pseudo}. In particular, $h ({\bf X})$ can be constructed from $H_{2\times 2}({\bf X})$ by the following similarity transformation,
\begin{eqnarray}
h({\bf X})=\eta({\bf X}) H_{2\times 2}({\bf X}) \eta^{-1}({\bf X}),
\label{smallh}
\end{eqnarray}
where the transformation matrix $\eta({\bf X})$ satisfies $\eta^{2}({\bf X})=W_{2\times 2}({\bf X})$.
One explicit form of $\eta({\bf X})$ is found to be \cite{qh}
\begin{equation}
\eta({\bf X})=\sqrt{\frac{\sqrt{a^2-b^2}}{2(|a|+ \sqrt{a^2-b^2})}}[W_{2\times 2}({\bf X})+ 1].
\label{eta}
\end{equation}
Equations (\ref{smallh}) and (\ref{eta}) lead to $
h({\bf X})= \varepsilon + {\cal U}  |a|\vec{e}_r\cdot \vec{\sigma}$.
It is noteworthy that $h({\bf X})$ is seen to be a standard Hermitian Hamiltonian describing a spin-$\half$ particle in a magnetic field.

Returning to \Eqn{eqn:Schrodinger} using the representation of $|\psi_{\eta}(t)\>\equiv \eta({\bf X})|\Psi(t)\>$, one finds
\begin{eqnarray}
i\frac{d}{dt} |\psi_{\eta}(t)\> &=& h[{\bf X}(t)] |\psi_{\eta}(t)\> + v [{\bf X}(t)] |\psi_{\eta}(t)\>,
\label{evolution2}
\end{eqnarray}
where the Hermitian potential function $v ({\bf X})$ is given by
\begin{eqnarray}
v ({\bf X})& = & i \eta({\bf X}) \left[ \eta^{-1}({\bf X}){\bm \nabla} \eta({\bf X})\right.  \nn \\
&&\ \left.-  {\bf M}_{2\times 2}({\bf X})\right] \eta^{-1}({\bf X})  \cdot \frac{d{\bf X}}{dt}.
\label{v}
\end{eqnarray}
 Note that if ${\bf M}_{2\times 2}({\bf X})=\eta^{-1}({\bf X}){\bm \nabla} \eta({\bf X})$, then $v({\bf X})=0$. In this special case the dynamical equation for $|\psi_{\eta}(t)\>$ is solely determined by the familiar Hamiltonian $h ({\bf X})$. This fact further confirms that the dynamics associated with a time-varying metric in \cP\cT-symmetric QM is a legitimate extension of the stationary-metric cases considered in Refs.~\cite{metric1,metric2}.

 Of more significance, because ${\bf M}_{2\times 2}({\bf X})$ can be chosen to be Hermitian [see Eq.~(\ref{m22})] and $\eta^{-1}({\bf X}){\bm \nabla} \eta({\bf X})$ is not Hermitian in general, we have $\eta^{-1}({\bf X}){\bm \nabla} \eta({\bf X})-{\bf M}_{2\times 2}({\bf X})\ne 0$.  As such, for  each choice of ${\bf M}_{2\times 2}({\bf X})$ satisfying \Eqn{M-eq2}, we have a different realization of $v ({\bf X})$ from \Eqn{v}. An interesting class of time-dependent problems in conventional QM can thus be defined from a $\cP\cT$-symmetric framework, always producing a Hermitian perturbative term dependent upon $d{\bf X}/dt$, the velocity in the parameter space.  Within the conventional QM, such type of perturbation may also arise from a designed feedback mechanism.  Alternatively, so long as a dynamical model (could be phenomenological, e.g., a system under a mixed quantum-classical description) requires a $d{\bf X}/dt$-dependent effective Hamiltonian, then a term similar to $v ({\bf X})$ emerges and all other terms in higher orders of $d{\bf X}/dt$ can be safely neglected in adiabatic processes.

 It is also evident why the nonzero $v ({\bf X})$ term can be important for adiabatic processes. In the adiabatic limit, though the perturbation $v ({\bf X})$ is vanishingly small, its effect will be accumulated over a diverging time scale. In brief, the expectation value of  $v ({\bf X})$ on the instantaneous eigenstate, which is of the order of $d{\bf X}/dt$, constantly corrects the quantum phase.  The accumulation of this phase correction gives a nonzero and finite extra phase of a geometric nature. This further justifies the emergence of a non-conventional Berry-like phase as well as a non-trivial $\gamma_{B}^{+}$ even at $\theta=0,\pi$.
  This ``Hermitian" perspective is also verified by our calculations based directly on \Eqn{evolution2}.  Indeed, due to the mapping from \Eqn{eqn:Schrodinger} to \Eqn{evolution2}, an adiabatic process for $|\Psi(t)\>$ is mapped to that for $|\psi_{\eta}(t)\>$, and the Berry-like phase for $|\Psi(t)\>$ is the same as that for $|\psi_{\eta}(t)\>$.

In studies of the adiabatic evolution associated with a Gross-Pitaeviski equation, recently Fu and Liu \cite{fuliu} also found a remarkable Berry-like phase.  In their model the instantaneous eigenstate changes adiabatically and at the same time induces a perturbation through a state-dependent mean-field Hamiltonian, with the perturbation strength considered to the first order of the velocity in the parameter space. This novel feedback mechanism clearly belongs to, and is also one explicit example of, the new class of adiabatic evolution found here. 

To conclude, adiabatic unitary evolution with a time-dependent metric is shown to exist in
\cP\cT-symmetric QM. For a \cP\cT-symmetric two-level system, a Berry-like phase is found to display unusual features when compared with current notion of the magnetic monopole and the Dirac string.  Our results establish an entire class of new geometric phase problems for both \cP\cT-symmetric QM and conventional QM, where adiabatic manipulation yields additional phase corrections to the conventional Berry-connection integral.
This work therefore represents another interesting application of  \cP\cT-symmetric QM with unexpected findings. J.G. is funded by NUS ``YIA" (WBS grant No. R-144-000-195-101). Helpful discussions with Jie Liu, Li-Bin Fu, and Qi Zhang are gratefully acknowledged.


\begin{thebibliography}{99}

\bibitem{BB}
C.M.~Bender and S.~Boettcher,
Phys.~Rev.~Lett.~{\bf 80}, 5243 (1998);
C.M.~Bender, D.C.~Brody, and H.F.~Jones,
{\it ibid}.~{\bf 89}, 270401 (2002).
\bibitem{Review}
C.M.~Bender,
Rep.~Prog.~Phys.~{\bf 70}, 947 (2007).
\bibitem{exp} A.~Guo {\it et al}., \prl{\bf 103}, 093902 (2009); C.E.~R\"{u}ter {\it et al.}, Nature Physics, doi: 10.1038/NPHYS1515 (2010).
\bibitem{theory}Z.H.~Musslimani, K.G.~Makris, R.~El-Ganainy, and  D.N.~Christodoulides, \prl{\bf 100}, 030402 (2008); K.G.~Makris, R.~El-Ganainy, D.N.~Christodoulides, and Z.H.~Musslimani, {\it ibid}.~{\bf 100}, 103904 (2008); S.~Longhi, {\it ibid.} {\bf 103}, 123601 (2009); C.T.~West, T.~Kottos, and T.~Prosen, {\it ibid}.~{\bf 104}, 054102 (2010).
\bibitem{Berry}M.V.~Berry, Proc.~R.~Soc.~Lond.~A {\bf 392}, 45 (1984).
\bibitem{book} 
For example, {\it Geometric Phases in Classical and Quantum Mechanics}, D.~Chru\'{s}ci\'{n}ski and A.~Jamiolkowski (Birkh\"{a}user, 2004).
\bibitem{qh}Q.-h.~Wang, S.-z.~Chia, and J.-h.~Zhang, ArXiv:1002.2676.
\bibitem{wright} J.G.~Garrison and E.M.~Wright, Phys.~Lett.~A {\bf 128}, 177 (1988).
\bibitem{pseudo}A.~Mostafazadeh, J.~Math.~Phys. {\bf 43}, 205 (2002); {\bf 43}, 2814 (2002).
\bibitem{metric2}C.~Figueira de Morrison and A.~Fring, J.~Phys.~A {\bf 39}, 9269 (2006).
\bibitem{metric1}A.~Mostafazadeh, Phys.~Lett.~B {\bf 650}, 208 (2007).
\bibitem{fuliu} L.B.~Fu and J.~Liu, arXiv:0906.3081; J.~Liu and L.B.~Fu, arXiv:0908.4130.
\end{thebibliography}
\end{document}